\documentclass[10pt,floatfix,aps,prl,amsmath,nofootinbib,preprintnumbers,twocolumn,preprint]{revtex4}
\usepackage{graphicx}
\usepackage{epstopdf}
\usepackage{amsmath}
\usepackage{amsfonts}
\usepackage{amssymb}
\usepackage{color}
\usepackage{mathrsfs}
\usepackage{multirow}
\usepackage{bm}
\usepackage{subfigure}
\usepackage{xcolor}
\usepackage{mathtext,bbm,amsmath,amsfonts,amssymb,indentfirst,syntonly,graphicx}
\usepackage{mathtools}
\usepackage{slashbox}
\usepackage[english]{babel}
\usepackage{calc}
\usepackage{tikz}

\def\({\left(}
\def\){\right)}
\def\[{\left[}
\def\]{\right]}
\def\e{\begin{equation}}
\def\q{\end{equation}}
\def\m{\begin{eqnarray}}
\def\n{\end{eqnarray}}

\begin{document}

\title{Polarization of gamma-ray burst afterglows in the synchrotron self-Compton process from a highly relativistic jet}

\author{Hai-Nan Lin$^{1}$}

\author{Xin Li$^{1,2}$}

\author{Zhe Chang$^{3}$}

\thanks{Corresponding author at: linhn@ihep.ac.cn}

\affiliation{$^1$Department of Physics, Chongqing University, Chongqing 401331, China\\
$^2$State Key Laboratory of Theoretical Physics, Institute of Theoretical Physics, Chinese Academy of Sciences, Beijing 100190, China\\
$^3$Institute of High Energy Physics, Chinese Academy of Sciences, Beijing 100049, China\\}

\date{\today}

\begin{abstract}
Linear polarization have been observed in both the prompt phase and afterglow of some bright gamma-ray bursts (GRBs). Polarization in the prompt phase spans a wide range, and may be as high as $\gtrsim 50\%$. In the afterglow phase, however, it is usually below $10\%$. According to the standard fireball model, GRBs are produced by synchrotron radiation and Compton scattering process in a highly relativistic jet ejected from the central engine. It is widely accepted that prompt emissions occur in the internal shock when shells with different velocities collide with each other, and the magnetic field advected by the jet from the central engine can be ordered in large scale. On the other hand, afterglows are often assumed to occur in the external shock when the jet collides with interstellar medium, and the magnetic field produced by the shock through, e.g. Weibel instability, is possibly random. In this paper, we calculate the polarization properties of the synchrotron self-Compton process from a highly relativistic jet, in which the magnetic field is randomly distributed in the shock plane. We also consider the generalized situation where a uniform magnetic component perpendicular to the shock plane is superposed on the random magnetic component. We show that, the polarization is hardly to be larger than $10\%$ if the seed electrons are isotropic in the jet frame. This may account for the observed upper limit of polarization in the afterglow phase of GRBs. In addition, if the random and uniform magnetic components decay with time in different speeds, then the polarization angle may change $90^{\circ}$ duration the temporal evolution.
\end{abstract}

\maketitle


\section{I. Introduction}\label{sec:introduction}

Gamma-ray bursts (GRBs) are one of the most violent explosions occurring in the deep universe. For recent reviews, see e.g. Ref.\cite{Piran:1999,Meszaros:2006,Kumar:2015}. Since their accidental discovery in 1967 by Vela satellites, significant progresses have been made in the past decades in understanding the properties of GRBs. Thanks to the combined contributions from various ground-based and space-borne instruments, the time resolved spectra and light curves can be observed with high precision in wide energy bands. However, several models can explain the observed spectra and light curves equally well. The emission mechanism of GRBs remains a mystery after about half a century researches. The polarimetric observations, on the other hand, provide a useful supplementary to the spectra and light curves to reveal the central engine.

Recently, it is found that photons in the prompt phase of some bright GRBs are highly linearly polarized, and may be as high as $\gtrsim 50\%$ \cite{Coburn:2003,McGlynn:2007,Kalemci:2007,Gotz:2009,Yonetoku:2011,Yonetoku:2012}. For example, Coburn \& Boggs \cite{Coburn:2003} claimed to have detected a polarization of $80\%\pm 20\%$ in GRB 021206, although a later reanalysis found no significant polarization signal in this same burst \cite{Rutledge:2004}. Kalemci et al. \cite{Kalemci:2007} analyzed the data of GRB 041219A and obtained a polarization fraction $98\%\pm 33\%$, in spite of the large statistical uncertainty. A detailed analysis of GRB 041219A showed that polarization degree is anti-correlated with photon energy \cite{McGlynn:2007}. The temporal evolution of polarization has also been detected \cite{Gotz:2009,Yonetoku:2011}. Interestingly, it was found with high confidence that the polarization angle of GRB 100826A changes $\sim 90^{\circ}$ between two time periods \cite{Yonetoku:2011}. Photons in the afterglow can also be linearly polarized. However, polarization in the afterglow is in general much smaller than that in the prompt phase, and it is usually below $10\%$. For example, Hjorth et al. \cite{Hjorth:1999} found an upper limit of $2.3\%$ on the polarization of optical afterglow of GRB 990123. Covino et al. \cite{Covino:2002} found the $3\sigma$ upper limit of $P < 2.7\%$ in the optical afterglow of GRB 011211. Bersier et al. \cite{Bersier:2003} reported a $9.9\%$ polarization in the optical afterglow of GRB 020405. Steele et al. \cite{Steele:2009} showed that the early optical afterglow of GRB 090102 was polarized at $10\%\pm 1\%$. Mundell et al. \cite{Mundell:2013} reported the detection of polarization $28\%\pm 4\%$ in the immediate afterglow of GRB 120308A, which decreases to $\sim 16\%$ over the subsequent ten minutes. Circular polarization is also possible, but it is very small, especially in the afterglow phase \cite{Matsumiya:2003a,Wiersema:2012,Wiersema:2014}.

Several theoretical models have been proposed to account for the polarimetric observations, see e.g. Ref.\cite{Covino:2016cuw} for recent review. One of the most promising mechanisms is the synchrotron radiation. It is well known that photons produced by synchrotron radiation can be highly polarized. The polarization properties of synchrotron radiation strongly depends on the magnetic configurations. There are three magnetic configurations that are widely discussed in literatures: (1) ordered magnetic field in the shock plane \cite{Granot:2003,Granot:2003b,Nakar:2003,Lan:2016a}, (2) ordered magnetic field perpendicular to the shock plane \cite{Granot:2003b}, and (3) random magnetic field confined in the shock plane \cite{Ghisellini:1999,Waxman:2003,Toma:2009,Lan:2016a}. The first two globally ordered magnetic fields can be carried out by the jet from the central engine \cite{Spruit:2001,Fendt:2004}, while the random magnetic field can be produced by the shock \cite{Gruzinov:1999a,Medvedev:1999}.

An alternative mechanism able to produce highly polarized photons is the Compton scattering process. Lazzati et al. \cite{Lazzati:2004} calculated the polarization properties of an isotropic photon field upscattered by a relativistic jet, and found that the polarization can be as large as in the point-source limit. However, they only discussed in the Thomson limit, and the seed photons were completely unpolarized. In fact, the seed photons are very likely to originate from synchrotron radiation, thus are initially polarized. Krawczynski \cite{Krawczynski:2012} calculated the polarization properties of inverse Compton emission and synchrotron self-Compton (SSC) emission using numerical simulations, and found that large polarization is possible. In a series of recent papers \cite{Chang:2013,Chang:2014a,Chang:2014b}, we presented a detailed calculation on the polarization properties of an initially polarized photon scattered by isotropic electrons with arbitrary energy distribution. We found that the final polarization spans a wide range, depending on the initial polarization state of the incident photon and the energy distribution of electrons. In this model, both the energy dependence of polarization degree in GRB 041219A and the change of polarization angle in GRB 100826A can be naturally explained \cite{Chang:2014d}.

The magnetic-dominated jet model is gradually becoming one of the most popular models nowadays due to its ability to explain a number of observational phonomania, such as spectra, light curves and polarization \cite{Meszaros2011,Veres:2012sb,Veres:2013rj}. More importantly, in the magnetic-dominated jet model, the jet can be effectively accelerated to a highly relativistic velocity, and the radiation efficiency can be large compared to the baryon-dominated jet model \cite{Spruit:2001,Drenkhahn2002,Drenkhahn2002b,Fendt:2004}. The SSC process is a natural result of the magnetic-dominated jet model. According to this model, GRB prompt emissions are produced through synchrotron radiation and Compton scattering process in a highly relativistic jet dominated by Poynting flux. The magnetic field ejected from the central engine is likely to be ordered in large scale, either parallel or perpendicular to the shock plane. Afterglows are assumed to be produce through synchrotron radiation in the external shock region when the jet collides with interstellar medium. The magnetic field produced by the shock in interstellar medium is possibly random. In the early afterglow region, the electron density may be still large enough such that the Compton scattering may play a role. In fact, Sari \& Esin \cite{Sari:2001} have computed the spectrum of the inverse Compton emission in afterglow and
found that it can dominate the total cooling rate of the afterglow for several months or even years after the prompt emission. The SSC radiation may occur in afterglow phase if the GRBs explode in a reasonably dense medium. In a recent paper \cite{Chang:2014c}, we have calculated the polarization properties of the SSC process from a highly relativistic jet in two ordered magnetic configurations. We found that in both cases, a maximum polarization degree of $\gtrsim 20\%$ is possible if the seed electrons are isotropic. We also showed that the polarization-luminosity relation in these two magnetic configurations is very different, which can be used to constrain the magnetic configuration in the future when a large amount of polarimetric data is available. There is another widely discussed magnetic configuration in literature but have not involved in our previous calculation, i.e., random magnetic field confined in the shock plane. It is useful to calculate the polarization properties in such a magnetic configuration, because GRB afterglows are often assumed to occur in external shock region where the magnetic field is likely to be random. In this paper we will give a detailed calculation on the polarization properties of the SSC process in random magnetic field.

The rest of the paper is organized as follows. In Section II, we calculate the power spectrum and polarization properties of the synchrotron radiation in random magnetic field. In Section III, we briefly review the polarization of a photon scattered by electrons with any spectral distribution. In section IV, we calculate the polarization properties of the SSC process from a highly relativistic jet, in which the magnetic field is randomly distributed in the shock plane. Finally, discussions and conclusions are given in Section V.

\section{II. Synchrotron radiation in random magnetic field}\label{sec:synchrotron}

In this section, we calculate the power spectrum and polarization properties of synchrotron radiation in random magnetic field case. We assume that the magnetic field is uniform in strength but random in direction, and it is fully confined in the shock plane. We further assume that the length scale of magnetic field is larger than the Lammor radius such that the classical synchrotron radiation formulae are applicable.

\begin{figure}[htbp]
\begin{center}
  \includegraphics[width=0.45\textwidth]{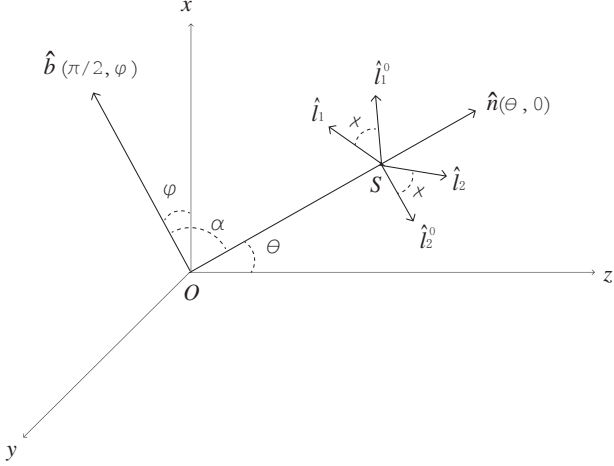}
  \caption{\small{Geometrical representation of the synchrotron radiation in random magnetic field confined in the $O\hat{\bm x}\hat{\bm y}$ plane. We choose a Cartesian coordinate such that the line-of-sight ($\hat{\bm n}$) is in the $O\hat{\bm x}\hat{\bm z}$ plane. $\hat{\bm b}$ is an arbitrary direction in the $O\hat{\bm x}\hat{\bm y}$ plane, representing the direction of magnetic field. $\hat{\bm l}_1$ and $\hat{\bm l}_2$ are two unit vectors in the $O\hat{\bm b}\hat{\bm n}$ plane and perpendicular to this plane, respectively, and $\hat{\bm n}=\hat{\bm l}_1\times \hat{\bm l}_2$. The $S\hat{\bm l}_1^0\hat{\bm l}_2^0$ frame is the $S\hat{\bm l}_1\hat{\bm l}_2$ frame rotating an angle $\chi$ with respect to $\hat{\bm n}$ direction, such that $\hat{\bm l}_1^0$ is in the $O\hat{\bm x}\hat{\bm z}$ plane, therefore $\hat{\bm l}_2^0$ is perpendicular to the $O\hat{\bm x}\hat{\bm z}$ plane.}}\label{fig:Geometry1}
\end{center}
\end{figure}

We first briefly review the properties of synchrotron radiation in uniform magnetic field. The geometry of synchrotron radiation is depicted in Fig.\ref{fig:Geometry1}. The magnetic field is along the $x$-axis, and the electron velocity is along the $\hat{\bm n}$ direction. The Cartesian coordinate is chosen such that $\hat{\bm n}$ is in the $O\hat{\bm x}\hat{\bm z}$ plane. $\hat{\bm l}_1^0$ and $\hat{\bm l}_2^0$ are two unit vectors in the $O\hat{\bm x}\hat{\bm z}$ plane and perpendicular to this plane, respectively, and $\hat{\bm n}=\hat{\bm l}_1^0\times \hat{\bm l}_2^0$. Recall that for a relativistic electron moving in the uniform magnetic field $\bm{B}$, the radiation is mainly confined in a small cone centering on the direction of electron velocity. The radiating power spectrum of a single electron is given by \cite{Rybicki:1979}
\begin{equation}\label{eq:power}
  P(\omega)=\frac{\sqrt{3}e^3B\sin\alpha}{2\pi m_e c^2}F\left(\frac{\omega}{\omega_c}\right),
\end{equation}
where $F(x)\equiv x\int_x^{\infty}K_{5/3}(\xi)d\xi$, $K_{5/3}(\xi)$ is the modified Bessel function of order $5/3$, $\omega_c\equiv 3\gamma_e^2eB\sin\alpha/2m_ec$ is the critical frequency, $\gamma_e$ is the Lorentz factor of the electron, $m_e$ is the rest mass of electron, and $\alpha$ is the pitch angle, i.e., the angle between the electron velocity and the magnetic field.

The radiating power spectrum can be divided into two parts of different polarization states, $P(\omega)=P_1(\omega)+P_2(\omega)$, where
\begin{eqnarray}
  \begin{cases}
    \displaystyle P_1(\omega)=\frac{\sqrt{3}e^3B\sin\alpha}{4\pi m_e c^2}\left[F\left(\frac{\omega}{\omega_c}\right)-G\left(\frac{\omega}{\omega_c}\right)\right],\\
    \displaystyle P_2(\omega)=\frac{\sqrt{3}e^3B\sin\alpha}{4\pi m_e c^2}\left[F\left(\frac{\omega}{\omega_c}\right)+G\left(\frac{\omega}{\omega_c}\right)\right].
  \end{cases}
\end{eqnarray}
Here $G(x)\equiv xK_{2/3}(x)$, where $K_{2/3}(\xi)$ is the modified Bessel function of order $2/3$. $P_1(\omega)$ and $P_2(\omega)$ represent the power polarized along $\hat{\bm l}_1^0$ and $\hat{\bm l}_2^0$ directions, respectively. The polarization degree is defined by
\begin{equation}\label{eq:pi_single}
  \Pi_{\rm syn}(\omega)=\frac{P_2(\omega)-P_1(\omega)}{P_2(\omega)+P_1(\omega)}.
\end{equation}

For power-law electrons, ${\mathcal N}(\gamma_e)d\gamma_e\propto\gamma_e^{-p}d\gamma_e$, we can integrate over the electrons to obtain the total radiating power. Through a straightforward calculation we obtain
\begin{eqnarray}\label{eq:power_PL}
  \begin{cases}
    \displaystyle P_1^{\rm PL}(\omega)\propto\left(\frac{p+7/3}{p+1}-1\right)(B\sin\alpha)^{\frac{p+1}{2}}\omega^{-\frac{p-1}{2}},\\
    \displaystyle P_2^{\rm PL}(\omega)\propto\left(\frac{p+7/3}{p+1}+1\right)(B\sin\alpha)^{\frac{p+1}{2}}\omega^{-\frac{p-1}{2}}.\\
  \end{cases}
\end{eqnarray}
The polarization degree is obtained from Eq.(\ref{eq:pi_single}) by the following replacements: $P_1(\omega)\rightarrow P_1^{\rm PL}(\omega)$, and $P_2(\omega)\rightarrow P_2^{\rm PL}(\omega)$. Thus we have
\begin{equation}\label{eq:pi_PL}
  \Pi_{\rm syn}^{\rm PL}=\frac{p+1}{p+7/3}.
\end{equation}
Note that the polarization degree is independent of photon energy, and it only depends on the power-law index of incident electrons. The index $p$ is usually in the range $2\lesssim p \lesssim 3$, thereby $\Pi_{\rm syn}^{\rm PL}\approx 70\%$ \cite{Sari:1999}.

Now consider that the magnetic field is along the direction of azimuth angle $\varphi$ in the $O\hat{\bm x}\hat{\bm y}$ plane, i.e., the $\hat{\bm b}$ direction in Fig.\ref{fig:Geometry1}. For power-law electrons moving in this magnetic field, the radiating power spectrum is also given by Eq.(\ref{eq:power_PL}). In this case, however, $P_1$ and $P_2$ are defined in a new frame $S\hat{\bm l}_1\hat{\bm l}_2$, which is the $S\hat{\bm l}_1^0\hat{\bm l}_2^0$ frame rotating an angle $\chi$ with respect to $\hat{\bm n}$ direction, such that $\hat{\bm l}_1$ is in the $O\hat{\bm b}\hat{\bm n}$ plane, and therefore $\hat{\bm l}_2$ is perpendicular to this plane. Rotating $P_1$ and $P_2$ back to the $S\hat{\bm l}_1^0\hat{\bm l}_2^0$ frame and averaging over the direction of magnetic field, we obtain
\begin{eqnarray}
  \begin{cases}
    \displaystyle\langle P_1(\omega)\rangle\propto\frac{1}{2\pi}\int_0^{2\pi}[P_1^{\rm PL}(\omega)\cos^2\chi+P_2^{\rm PL}(\omega)\sin^2\chi]d\varphi,\\
    \displaystyle\langle P_2(\omega)\rangle\propto\frac{1}{2\pi}\int_0^{2\pi}[P_1^{\rm PL}(\omega)\sin^2\chi+P_2^{\rm PL}(\omega)\cos^2\chi]d\varphi.\\
  \end{cases}
\end{eqnarray}
The squares of $\sin\chi$ and $\cos\chi$ arise from the fact that the radiating power is proportional to the square of the electric component of a photon. From geometrical consideration, we have
\begin{equation}\label{eq:angle1}
  \cos\alpha=\sin\theta\cos\varphi,~~ \cos\chi=\frac{\cos\theta\cos\varphi}{\sqrt{1-\sin^2\theta\cos^2\varphi}}.
\end{equation}
The total radiating power see at the viewing angle $\theta$ is given by the summation of $\langle P_1(\omega)\rangle$ and $\langle P_2(\omega)\rangle$,
\begin{equation}\label{eq:power_total}
  \langle P(\omega)\rangle\propto\frac{2(p+7/3)}{p+1}\frac{1}{2\pi}\int_0^{2\pi}(B\sin\alpha)^{\frac{p+1}{2}}d\varphi~\omega^{-\frac{p-1}{2}}.
\end{equation}
At the special viewing angle $\theta=0$, the total radiating power is equivalent to that when the magnetic field is uniformly distributed in the shock plane. In the special case $p=3$, Eq.(\ref{eq:power_total}) can be analytically integrated,
\begin{equation}
  \langle P(\omega)\rangle\propto\frac{2}{3}B^2(3+\cos2\theta)\omega^{-1}.
\end{equation}

Replacing $P_1(\omega)$ with $\langle P_1(\omega)\rangle$, and $P_2(\omega)$ with $\langle P_2(\omega)\rangle$ in Eq.(\ref{eq:pi_single}), we obtain the polarization of synchrotron radiation in random magnetic field,
\begin{eqnarray}\label{eq:pi_rand}\nonumber
  \langle\Pi_{\rm syn}(\theta)\rangle&=&\frac{\langle P_2(\omega)\rangle-\langle P_1(\omega)\rangle}{\langle P_2(\omega)\rangle+\langle P_1(\omega)\rangle}\\
   &=& \frac{p+1}{p+7/3}\frac{\int_0^{2\pi}(\sin\alpha)^\frac{p+1}{2}\cos2\chi d\varphi}{\int_0^{2\pi}(\sin\alpha)^\frac{p+1}{2} d\varphi}.
\end{eqnarray}
Note that the polarization degree is also independent of photon energy, and it only depends the viewing angle and the power-law index of incident electrons. When $\theta=\pi/2$, Eq.(\ref{eq:pi_rand}) reduces to Eq.(\ref{eq:pi_PL}) up to a minus sign, and it seems as if the magnetic field is globally ordered. In the special case $p=3$, Eq.(\ref{eq:pi_rand}) can be integrated analytically, leading to the result
\begin{equation}
  \langle\Pi_{\rm syn}(\theta)\rangle=-\frac{3}{2}\frac{\sin^2\theta}{3+\cos2\theta}.
\end{equation}
The minus sign means that photons are polarized along the $\hat{\bm l}_1^0$ direction.

In a more general case, there may be a uniform magnetic field component $\tilde{\bm B}$ perpendicular to the shock plane. The spectrum of synchrotron radiation from power-law electrons in this uniform magnetic field component can be derived from Eq.(\ref{eq:power_PL}) by replacing $B\sin\alpha$ with $\tilde{B}\sin\theta$. Therefore, the total radiating power in the composite magnetic fields ${\bm B}+\tilde{\bm B}$, seeing at the viewing angle $\theta$, is given by
\begin{eqnarray}\nonumber\label{eq:power_total2}
  \langle P_{\rm t}(\omega)\rangle\propto\frac{2(p+7/3)}{p+1}\bigg{[}\frac{1}{2\pi}\int_0^{2\pi}(B\sin\alpha)^{\frac{p+1}{2}}d\varphi \\ +(\tilde{B}\sin\theta)^{\frac{p+1}{2}}\bigg{]}\omega^{-\frac{p-1}{2}}.
\end{eqnarray}
The polarization degree of synchrotron radiation in the composite magnetic fields is given by
\begin{eqnarray}\label{eq:PI_composite}
  \langle\Pi_{\rm t}(\theta)\rangle=\frac{p+1}{p+7/3}\frac{\frac{1}{2\pi}\int_0^{2\pi}(\sin\alpha)^{\frac{p+1}{2}}\cos2\chi d\varphi+(R\sin\theta)^{\frac{p+1}{2}} }{\frac{1}{2\pi}\int_0^{2\pi}(\sin\alpha)^{\frac{p+1}{2}} d\varphi+(R\sin\theta)^{\frac{p+1}{2}}},
\end{eqnarray}
where $R\equiv\tilde{B}/B$ is the ratio between the uniform and random magnetic field components. When $p=3$, Eq.(\ref{eq:PI_composite}) reduces to
\begin{equation}
  \langle\Pi_{\rm t}(\theta)\rangle=-\frac{3}{2}\frac{(1-2R^2)\sin^2\theta}{(3+\cos2\theta)+4R^2\sin^2\theta}.
\end{equation}

\begin{figure}[htbp]
\begin{center}
  \includegraphics[width=0.5\textwidth]{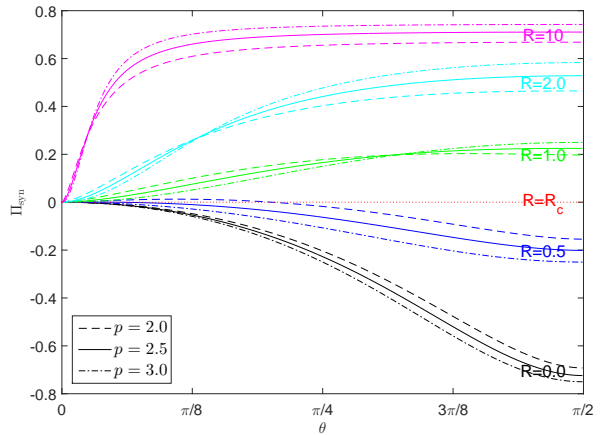}
  \caption{\small{(color online) The polarization degree of synchrotron radiation as a function of viewing angle for different values of $R$ ($R=0,0.5,1,2,10$) and $p$ ($p=2.0,2.5,3.0$) in the composite magnetic fields.}}\label{fig:PI_syn}
\end{center}
\end{figure}

\begin{figure}[htbp]
\begin{center}
  \includegraphics[width=0.5\textwidth]{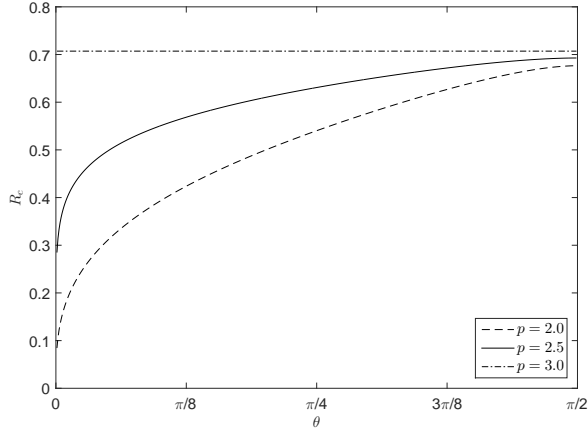}
  \caption{\small{The critical value $R_c$ as a function of $\theta$ for different values of $p$ ($p=2.0,2.5,3.0$) in the composite magnetic fields.}}\label{fig:Rc}
\end{center}
\end{figure}

We plot the polarization degree as a function of viewing angle for different values of $R$ ($R=0,0.5,1,2,10$) in Fig.\ref{fig:PI_syn}. For each $R$, we choose three different values of $p$, i.e. $p=2.0,2.5,3.0$. Curves for different $R$ are distinguished by color, and curves for different $p$ are distinguished by line style. We just plot in the $\theta\in[0,\pi/2]$ range because the polarization is symmetric with respect to $\theta=\pi/2$. From Fig.\ref{fig:PI_syn}, we can see that the polarization as a function of $\theta$ peaks at $\theta = \pi/2$. The polarization is insensitive to $p$, but it strongly depends on $R$. As $R$ increases from 0, the net polarization will firstly decreases. This is because the uniform magnetic field is perpendicular to the random magnetic field. The synchrotron radiations in these two magnetic field components are polarized along two directions which are perpendicular to each other. For the random magnetic field component, the radiation is polarized along $\hat{\bm l}_1^0$ direction, while for the uniform magnetic field component, it is polarized along $\hat{\bm l}_2^0$ direction. Therefore the net polarization is cancelled out. When $R$ reaches to a critical value $R_c$, the polarization completely vanishes. The concrete value of $R_c$ can be obtained by requiring that right-hand-side of Eq.(\ref{eq:PI_composite}) to be zero, and it depends on both $\theta$ and $p$. We plot $R_c$ as a function of $\theta$ for different values of $p$ ($p=2.0,2.5,3.0$) in Fig.\ref{fig:Rc}. In the specific case $p=3$, $R_c\equiv 1/\sqrt{2}$. When $R>R_c$, the uniform component dominates over the random component, and the net polarization degree begins to increase as $R$ increases. However, the polarization angle rotates $90^{\circ}$ with respect to the $R<R_c$ case. When $R\rightarrow \infty$, the random component is negligible, and Eq.(\ref{eq:PI_composite}) reduces to Eq.(\ref{eq:pi_PL}). In this case, the polarization is independent of viewing angle, and it seems as if the magnetic field is uniform in large scale.

\section{III. Compton scattering process}\label{sec:compton}

In this section, we shortly review the polarization properties of the Compton scattering process. The details can be found in Ref.\cite{Chang:2014a,Chang:2014b}. We first calculate the single scattering case in which a photon is scattered by an electron with arbitrary momentum. The incident photon can be in arbitrary polarization state. Then we integrate over the electrons to obtain the polarization of a photon scattered by isotropic electrons. Traditionally, the calculation is first done in the electron-rest frame, then is transformed to the laboratory frame. To avoid the complex Lorentz transformation between these two frames, we directly work in the laboratory frame.

\begin{figure}[htbp]
\begin{center}
  \includegraphics[width=0.5\textwidth]{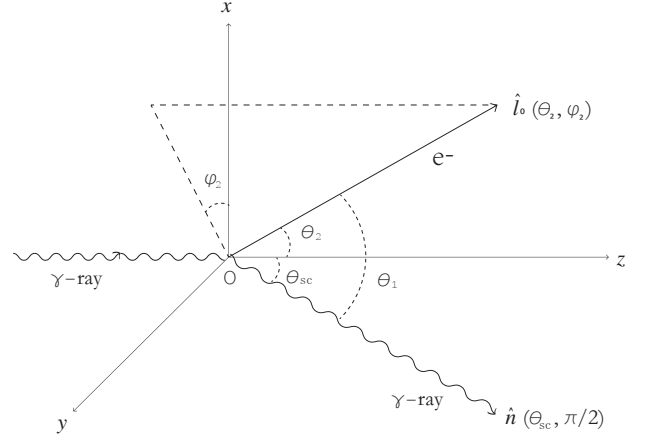}
  \caption{\small{Geometrical representation of the Compton scattering process in the laboratory frame. The incident photon initially goes along the positive $z$-axis, then is scattered by an electron moving along the $\hat{{\bm l}}_0$ direction at point $O$. After that the photon travels alone the line-of-sight direction $\hat{{\bm n}}$. We choose a Cartesian coordinate such that the $y$-axis is in the scattering plane.}}\label{fig:Geometry2}
\end{center}
\end{figure}

We first consider the single scattering case. The geometry of Compton scattering process is illustrated in Fig.\ref{fig:Geometry2}. A photon with energy $\varepsilon_0$ collides with an electron traveling along arbitrary direction $\hat{\bm l}_0$ at point $O$. After that the photon is scattered to the $\hat{\bm n}$ direction. The Lorentz factor of the electron is $\gamma_e$. We set a Cartesian coordinate such that the $z$-axis is along the direction of incident photon, and the $y$-axis is in the scattering plane. The polar and azimuth angles of the incident electron are denoted by $\theta_2$ and $\varphi_2$, respectively. The scattering angle is denoted by $\theta_{\rm sc}$, and the angle between $\hat{\bm l}_0$ and $\hat{\bm n}$ is denoted by $\theta_1$. The energy of scattered photon can be obtained from the conservation of the energy and momentum \cite{Akhiezer:1965},
\begin{equation}\label{eq:energy}
 \varepsilon_1=\frac{\varepsilon_0(1-\beta_e \cos\theta_2)}{\frac{\varepsilon_0}{\gamma_e m_e c^2}(1-\cos\theta_{\rm sc})+(1-\beta_e\cos\theta_1)},
\end{equation}
where $\beta_e=\sqrt{1-1/\gamma_e^2}$\, is the velocity of the incident electron in unit of light speed.

The polarization-dependent differential cross section of Compton scattering process in the laboratory frame, is given by \cite{Berest:1982}
\begin{eqnarray}\label{eq:cross-section}\nonumber
  d \sigma = \frac{1}{4} r_e^2 d \Omega \left(\frac{\varepsilon_1}{\varepsilon_0}\right)^2 [F_0 +F_3(\xi_3+\xi'_3) \\
   + F_{11} \xi_1 \xi_1' +F_{22} \xi_2\xi'_2+F_{33} \xi_3\xi'_3],
\end{eqnarray}
where $r_e=e^2/m_ec^2$ is the classical electron radius, $\xi_i$ and $\xi'_i$ ($i=1,2,3$) are the Stokes parameters standing for the polarization states of the incident and scattered photons, respectively. $\xi_3$ represents the linear polarization along $x$- or $y$-axis, $\xi_1$ represents the linear polarization along the direction with azimuth angle $\pm \pi/4$ with respect to the $x$-axis, and $\xi_2$ characterizes the circular polarization. The quantities $F_a$ $(a=0,3,11,22,33)$ depend on the kinematic states of the photon and electron, which are given by \cite{Chang:2014a,Chang:2014b}
\begin{eqnarray}\label{eq:fs}
 \begin{cases}
  F_0(\varepsilon_0;\gamma_e,\theta_2,\varphi_2;\theta_{\rm sc}) = \frac{\varepsilon_1}{\varepsilon_0}+\frac{\varepsilon_0}{\varepsilon_1}-\sin^2 \theta_{\rm sc},\\
  F_3(\gamma_e,\theta_2,\varphi_2;\theta_{\rm sc}) = -(A^2+A) \Sigma,\\
  F_{11}(\gamma_e,\theta_2,\varphi_2;\theta_{\rm sc}) = (A+\frac{1}{2}) \Sigma,\\
  F_{22}(\varepsilon_0;\gamma_e,\theta_2,\varphi_2;\theta_{\rm sc}) = \frac{1}{4}(1+2A)B \Sigma,\\
  F_{33}(\gamma_e,\theta_2,\varphi_2;\theta_{\rm sc}) = (A^2+A+\frac{1}{2}) \Sigma,
 \end{cases}
\end{eqnarray}
where
\begin{equation}\label{eq:ab}
  A \equiv \frac{1}{x}-\frac{1}{y}, ~ B\equiv \frac{x}{y}+\frac{y}{x},
\end{equation}
\begin{equation} \label{eq:xy}
  x\equiv\frac{2\gamma_e \varepsilon_0}{m_e c^2}(1-\beta_e \cos \theta_2), ~ y\equiv\frac{2\gamma_e \varepsilon_1}{m_e c^2}(1-\beta_e \cos \theta_1),
\end{equation}
\begin{equation}\label{eq:Sigma}
 \Sigma\equiv\frac{4}{\gamma_e^2(1-\beta_e\cos\theta_2)^2}\left(1-\frac{\beta_e\sin\theta_2\sin\varphi_2}{1-\beta_e\cos\theta_2} \tan\frac{\theta_{\rm sc}}{2}\right).
\end{equation}
In Eq.(\ref{eq:fs}), we have explicitly written the arguments of $F_a$ for clarity. Note that $F_3$, $F_{11}$ and $F_{33}$ are independent of photon energy.

The polarimetric observation of GRB in both the prompt and afterglow phases shows that the circular polarization is very small. Hence we ignore it in the following calculation. For a photon with linear polarization degree $\Pi_0$, we can write the Stokes parameters as
\begin{equation}
  \xi_1=\Pi_0\sin 2\chi_0,~~\xi_2=0,~~\xi_3=\Pi_0\cos 2\chi_0,
\end{equation}
where the polarization angle $\chi_0\in[-\pi/2,\pi/2]$ is the angle between the polarization vector and the $x$-axis. After the Compton scattering process, the Stokes parameters of the secondary photon are given by \cite{Berest:1982}
\begin{equation}\label{eq:stokes}
 \xi^{\rm f}_1=\frac{ \xi_1 F_{11}}{F_0+\xi_3 F_3}, \quad
 \xi^{\rm f}_2=\frac{ \xi_2 F_{22}}{F_0+\xi_3 F_3}, \quad
 \xi^{\rm f}_3=\frac{F_3+ \xi_3 F_{33}}{F_0+\xi_3 F_3}.
\end{equation}
As can been see, the secondary photon is circularly polarized if and only if the incident photon is circularly polarized. The polarization degree of the scattered photon can be conveniently written in terms of the Stokes parameters as
\begin{equation}\label{eq:pi0}
 \Pi=\sqrt{(\xi_1^{\rm f})^2+(\xi_2^{\rm f})^2+(\xi_3^{\rm f})^2}.
\end{equation}

If a photon is scattered by isotropic electrons whose energies follow the power-law distribution ${\mathcal N}_e(\gamma_e)d\gamma_e \propto \gamma_e^{-p}d\gamma_e$, we can integrate over the electrons to obtain the average contribution from each electron \cite{Chang:2014a,Chang:2014b},
\begin{equation}\label{eq:Fa_averaged}
 \langle F_a (\varepsilon_0,\theta_{\rm sc})\rangle\equiv\frac{1}{C}\int\left(\frac{\varepsilon_1}{\varepsilon_0}\right)^2 F_a{\cal N}_e(\gamma_e) d\gamma_e d\Omega_2,
\end{equation}
where $C\equiv\int{\cal N}_e(\gamma_e) d\gamma_e d\Omega_2$ is the normalization factor, and $d\Omega_2\equiv\sin\theta_2d\theta_2d\varphi_2$. The term $(\varepsilon_1/\varepsilon_0)^2$ in the integrand on the right-hand-side of Eq.(\ref{eq:Fa_averaged}) arises from the average of cross section in Eq.(\ref{eq:cross-section}). The Stokes parameters of the scattered photon can be derived from Eq.(\ref{eq:stokes}) by replacing $F_a$ with $\langle F_a\rangle$. The polarization of the scattered photon is given by
\begin{equation}\label{eq:pi-scattered}
 \langle\Pi(\varepsilon_0,\theta_{\rm sc})\rangle=\sqrt{\langle\xi^{\rm f}_1\rangle^2+\langle\xi^{\rm f}_2\rangle^2+\langle\xi^{\rm f}_3\rangle^2}.
\end{equation}

In the Thomson limit, $\varepsilon_0\ll m_ec^2$, the formulae can be extensively simplified. The polarization of the scattered photon can be simply written as \cite{Chang:2014b}
\begin{equation}\label{eq:pi-thomson}
  \langle\Pi(\theta_{\rm sc})\rangle=\Pi_0\frac{\langle F_{11} \rangle}{\langle F_0 \rangle},
\end{equation}
where
\begin{equation}
  F_0=\frac{\varepsilon_1}{\varepsilon_0}+\frac{\varepsilon_0}{\varepsilon_1}-\sin^2\theta_{\rm sc}, \frac{\varepsilon_1}{\varepsilon_0}=\frac{1-\beta_e \cos\theta_2}{1-\beta_e\cos\theta_1}, F_{11}=\frac{\Sigma}{2}.
\end{equation}
A initially unpolarized photon remains unpolarized after scattering. The polarization of the scattered photon is independent of photon energy.

\section{IV. SSC process from a highly relativistic jet}\label{sec:syn-compton}

\begin{figure}[htbp]
\begin{center}
  \includegraphics[width=0.45\textwidth]{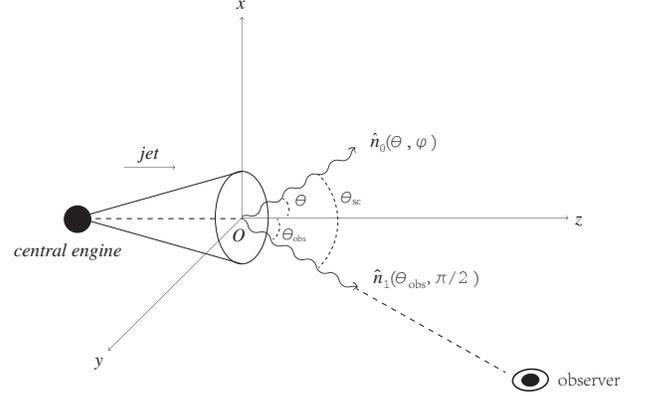}
  \caption{\small{Geometrical representation of the SSC process from a highly relativistic jet. A photon produced by synchrotron radiation is initially moving in the $\hat{\bm n}_0$ direction, then is scatted by isotropic electrons to the observer in $\hat{\bm n}_1$ direction.  We choose a Cartesian coordinate such that the $z$-axis is along the jet direction, and the observer is in the $O\hat{\bm y}\hat{\bm z}$ plane. The magnetic field is random in the $O\hat{\bm x}\hat{\bm y}$ plane.}}\label{fig:Geometry3}
\end{center}
\end{figure}

In this section, we calculate the polarization properties of the SSC process from a highly relativistic jet. The geometry is briefly illustrated in Fig.\ref{fig:Geometry3}. A highly relativistic jet ejected from the central engine collides with interstellar medium and produces external shocks. The shocks magnify magnetic field and, at the same time accelerate the electrons. Electrons moving in the magnetic field produce photons through synchrotron radiation. Then the synchrotron photons are scattered by the seed electrons through Compton scattering process.

The energy of electrons, as is predicted by the shock acceleration mechanisms, follows the power-law distribution. The magnetic field produced by shocks through Weibel instability is possibly random. We assume that electrons are isotropic in the jet frame, and the magnetic field is fully confined in the shock plane. We set a Cartesian coordinate such that the $z$-axis is in the jet direction, and therefore the magnetic field is in the $O\hat{\bm x}\hat{\bm y}$ plane. We also allow the existence of a uniform magnetic component perpendicular to the shock plane (i.e. along the $z$-axis). The radiation is axis-symmetric with respect to the $z$-axis. For simplicity we assume that the observer is in the $O\hat{\bm y}\hat{\bm z}$ plane. In such a configuration, the photon spectrum produced by synchrotron radiation, according to Eq.(\ref{eq:power_total2}), is given by
\begin{equation}
  N_{\gamma}(\varepsilon_0,\theta)\propto\left[\frac{1}{2\pi}\int_0^{2\pi}(\sin\alpha)^{\frac{p+1}{2}}d\varphi+(R\sin\theta)^{\frac{p+1}{2}}\right]\varepsilon_0^{-\frac{p-1}{2}},
\end{equation}
where $\theta$ is the angle with respect to the $z$-axis, $\alpha$ is given by Eq.(\ref{eq:angle1}), and $p$ is the power-law index of electrons. The photon initially moving in $\hat{\bm n}_0$ direction is scattered by seed electrons to observer in $\hat{\bm n}_1$ direction. The polarization degree of this photon seen by an observer is given by Eq.(\ref{eq:pi-scattered}), or in the Thomson limit given by Eq.(\ref{eq:pi-thomson}), where
\begin{equation}
  \cos\theta_{\rm sc}=\sin\theta\sin\varphi\sin\theta_{\rm obs}+\cos\theta\cos\theta_{\rm obs}
\end{equation}
is the scattering angle, and $\theta_{\rm obs}$ is the viewing angle, see Fig. \ref{fig:Geometry3}. Integrating over the photon spectrum, we obtain the polarization of the SSC process as a function of viewing angle,
\begin{equation}\label{eq:pi_ssc}
  \langle\langle\Pi(\theta_{\rm obs})\rangle\rangle=\frac{\int\langle\Pi(\varepsilon_0,\theta_{\rm sc})\rangle N_{\gamma} (\varepsilon_0,\theta) \sin\theta d\theta d\varphi d\varepsilon_0}{\int N_{\gamma} (\varepsilon_0,\theta) \sin\theta d\theta d\varphi d\varepsilon_0}.
\end{equation}

\begin{figure}[htbp]
\begin{center}
  \includegraphics[width=0.5\textwidth]{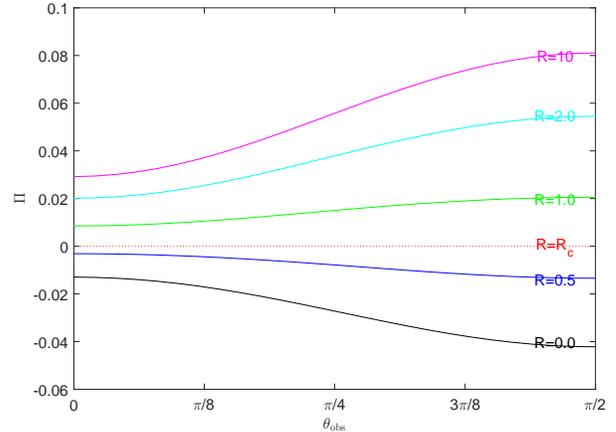}
  \caption{\small{(color online) Polarization of the SSC process in the Thomson limit. The Lorentz factors of the seed electrons are taken to be in the range $\gamma_e\in[1,10]$, and the power-law index of electrons is taken to be $p=2.5$.}}\label{fig:PI_ssc}
\end{center}
\end{figure}

The polarization measurement in GRB afterglow is usually performed in the optical band, in which the Thomson limit is applicable. Therefore, to simplicity we just have to calculate in the Thomson limit. We numerically integrate Eq.(\ref{eq:pi_ssc}), and plot the polarization as a function of viewing angle in Fig.\ref{fig:PI_ssc}. The figure is symmetric with respect to $\theta_{\rm obs}=\pi/2$ so we just plot in the $[0,\pi/2]$ range. Curves for different values of $R$ are shown. In the numerical calculation, the Lorentz factors of the seed electrons are taken to be in the range $\gamma_e\in[1,10]$, and the power-law index of electrons is taken to be $p=2.5$. Electrons with Lorentz factor larger than $10$ have little contribution to the scattering process \cite{Chang:2014a,Chang:2014b}. A different $p$ value does not significantly affect the results. Due to the isotropy of seed electrons and the randomness of magnetic field, the net polarization degree is highly suppressed. The polarization degree is hardly to be larger than $10\%$, unless the uniform magnetic component is at least ten times stronger than the random component such that the latter is negligible. The polarization degree increases with the increasing of viewing angle and peaks at $\theta_{\rm obs}\sim\pi/2$. When the random component is in the same order of magnitude with the uniform component, the net polarization almost vanishes. The polarization angle rotates $90^{\circ}$ when the magnetic field changes from random component dominated to uniform component dominated.

The above formulae are derived in the jet comoving frame. The jet travels towards the observer with a large bulk Lorentz factor $\Gamma$. The angle transformation between the jet frame and the observer frame is given by
\begin{equation}\label{eq:theta}
  \cos\theta_{\rm obs}=\frac{\cos\bar{\theta}_{\rm obs}-\beta_{\rm jet}}{1-\beta_{\rm jet}\cos\bar{\theta}_{\rm obs}},
\end{equation}
where $\beta_{\rm jet}=(1-1/\Gamma^2)^{1/2}$ is the velocity of the jet in unit of light speed. The quantities in the observer frame are denoted with a bar. Notice that the polarization is Lorentz invariant \cite{Cocke:1972}, we can easily transform the polarization from the jet frame to the observer frame, i.e.,
\begin{equation}\label{eq:Pi-obs}
  \langle\langle\bar{\Pi}(\bar{\theta}_{\rm obs})\rangle\rangle=\langle\langle\Pi(\theta_{\rm obs})\rangle\rangle.
\end{equation}
The net polarization degree as a function of viewing angle in the observer frame is very similar to Fig.\ref{fig:PI_ssc}, except that the $x$-axis is rescaled according to Eq.(\ref{eq:theta}).

\section{V. Discussions and conclusions}\label{sec:conclusion}

In this paper, we presented a detailed calculation on the polarization properties of the the SSC process from a highly relativistic jet in random magnetic field case. We assumed that the magnetic field is confined in a plane perpendicular to the jet velocity. This magnetic configuration is physically important and has been extensively discussed in literatures. Such a magnetic configuration may be produce by external shock through Weibel instability, while GRB afterglows are widely accepted to be produced in the external shock region. In addition to the random magnetic component, there may be a uniform component perpendicular to the shock plane. This uniform magnetic field can be advected by jet from the central engine. We first derived analytical formulae to calculate the power spectrum and polarization of synchrotron radiation in the composite magnetic fields. Starting from the polarization-dependent differential cross section of photon-electron scattering, we obtained the polarization of a photon scattered by an electron. Then integrating over the spectra of photons and seed electrons, the polarization properties of the SSC process were derived. We numerically calculated the polarization degree as a function of viewing angle in the Thomson limit. We found that the maximum polarization degree is usually $\lesssim 10\%$ if the seed electrons are isotropically distributed. This is consistent with the observation fact that the polarization of GRB afterglow is seldom larger than $10\%$.

Interestingly, if the two magnetic components (random and uniform) decay with time in different speeds, then the polarization angle may rotate $90^{\circ}$ during the temporal evolution. For example, the flux conservation requires that the transverse (random) component decays as $B\propto r^{-1}$, and the radial (uniform) component decays as $\tilde{B}\propto r^{-2}$, where $r$ is the distance to the central engine \cite{Spruit:2001}. At first, the uniform component dominates over the random component, i.e. $R\equiv\tilde{B}/B\gg 1$. As the jet expands, both components decay. Since the uniform component decays faster than the random component, after a critical radius the latter will dominate over the former and thus $R\ll 1$. The polarization angle will change $90^{\circ}$ at a critical value $R=R_c$.

Note that the polarization measurements of GRB afterglow are usually performed in the optical band. In this low energy band, the cross-section of photon-electron collision is small and therefore the Compton scattering process is negligible. The Compton scattering is important in the gamma-ray and $X$-ray band, and the synchrotron radiation may be the main radiating mechanism of optical afterglow. However, Compton scattering can still play an important role in the optical afterglow if the surrounding medium is dense \cite{Sari:2001}. Even if the Compton scattering does not occur, the $90^{\circ}$ change of polarization angle can also happen if the two magnetic components decay in different speed. The polarization degree of pure synchrotron radiation can vary in a wide range, depending on the ratio between the uniform and random magnetic components. The effect of Compton scattering is to suppress the polarization degree.

\vspace{4mm}

\begin{acknowledgments}
We are grateful to Y. Sang, P. Wang and Z. C. Zhao for useful discussions. This work has been supported by the Fundamental Research Funds for the Central Universities (Grant No. 106112016CDJCR301206), the National Natural Science Fund of China (Grant Nos. 11375203 and 11603005), and the Open Project Program of State Key Laboratory of Theoretical Physics, Institute of Theoretical Physics, Chinese Academy of Sciences, China (Grant No. Y5KF181CJ1).
\end{acknowledgments}



\end{document}